\documentclass[submission,copyright]{eptcs}
\providecommand{\event}{TTC 2011} 

\usepackage{graphicx} 
\usepackage[utf8]{inputenc}
\usepackage{url}
\usepackage{amsmath}
\usepackage{amsfonts}
\usepackage{amssymb}
\usepackage{subfigure}
\usepackage{wrapfig}
\usepackage{tabularx}
\usepackage{cite}

\usepackage{color}
\definecolor{listinggray}{rgb}{0.9,0.9,0.9}
\definecolor{keywordcolor}{rgb}{0.5,0,0.1}
\definecolor{commentcolor}{rgb}{0,0.5,0}
\definecolor{stringcolor}{rgb}{0,0,1}
\usepackage[hyper]{listings}

\lstdefinelanguage{viatra}
{morekeywords={@Trigger,trigger,shareable,guard,asmfunction,rule,gtrule,if,do,choose,forall,iterate,print,println,log,apply,
    entity,relation,supertypeOf,subtypeOf,typeOf,instanceOf,try,else,pattern,
    precondition,postcondition,action,neg,find,import,namespace,in,below,out,
    inout,let,multiplicity,many_to_one,many_to_many,one_to_many,one_to_one,
    isAggregation,inverse,seq,update,ref,true,false,call,machine,or,
    undef,rename,new,del,delete,move,copy,setValue,setFrom,setTo,with,when,check,change,appear,disappear,upon,cdrule},
 sensitive=true,
 morecomment=[l]{//},
 morecomment=[s]{/*}{*/},
 morestring=[b]{"},
}

\newcommand{\callLink}[1]{\hyperlink{#1}{#1}}
\newcommand{\callQualifiedLink}[2]{\hyperlink{#1}{#2}}
\newcommand{\callAnchor}[1]{\hypertarget{#1}{#1}}
\newcommand{\callQualifiedAnchor}[2]{\hypertarget{#1}{#2}}

\newcommand{\Viatra}{\textsc{Viatra2}}

\newcommand{\code}[1]{\lstinline[basicstyle=\small\ttfamily]!#1!}

\begin{document}

\title{Saying Hello World with \Viatra{} -\\ A Solution to the TTC 2011 Instructive Case\thanks{This work was partially supported by ICT FP7 SecureChange (ICT-FET-231101) European Project.}}
\author{{\'A}bel Heged{\"u}s \qquad\qquad Zolt\'an Ujhelyi \qquad\qquad G\'abor Bergmann
\institute{
  Fault Tolerant Systems Research Group \\
  Department of Measurement and Information Systems \\
  Budapest University of Technology and Economics, Hungary \\
  \email{\quad hegedusa@mit.bme.hu \quad\qquad ujhelyiz@mit.bme.hu \quad\qquad bergmann@mit.bme.hu}
  }
}

\def\titlerunning{Saying Hello World with \Viatra{} - A Solution to the TTC 2011 Instructive Case}
\def\authorrunning{{\'A}. Heged{\"u}s, Z. Ujhelyi \& G. Bergmann}

\maketitle

\begin{abstract}
The paper presents a solution of the \emph{Hello World! An Instructive Case for the Transformation Tool Contest} 
 using the \Viatra{} model transformation tool.

\end{abstract}
%

\section{Introduction}
Automated model transformations play an important role in modern model-driven system engineering in order to query, derive and manipulate large, industrial models. Since such transformations are frequently integrated into design environments, they need to provide short reaction time to support software engineers.

The objective of the \Viatra~(VIsual Automated model TRAnsformations~\cite{viatra}) framework is to support the entire life-cycle of model transformations consisting of specification, design, execution, validation and maintenance. 

\emph{Model representation.} \Viatra\ uses the VPM (Visual and Precise Metamodeling) approach~\cite{sosym2003_vpm} for describing modeling languages and models. The main reason for selecting VPM instead of a MOF-based metamodeling approach is that VPM supports arbitrary metalevels in the model space. As a direct consequence, models taken from conceptually different domains (and/or technological spaces) can be easily integrated into the VPM model space. The flexibility of VPM is demonstrated by a large number of already existing model importers accepting the models of different BPM formalisms, UML models of various tools, XSD descriptions, and EMF models.

\emph{Graph transformation} (GT) \cite{GT:HandbookII} based tools have been frequently used for specifying and executing complex model transformations. In GT tools, \emph{graph patterns} capture structural conditions and type constraints in a compact visual way. At execution time, these conditions need to be evaluated by \emph{graph pattern matching}, which aims to retrieve one or all matches of a given pattern  to execute a transformation rule. A \emph{graph transformation rule} declaratively specifies a model manipulation operation, that replaces a match of the LHS (left-hand side) graph pattern with an image of the RHS (right-hand side) pattern.

\emph{Transformation description.} Specification of model transformations in \Viatra\ combines the visual, declarative rule and pattern based paradigm of graph transformation and the very general, high-level formal paradigm of abstract state machines (ASM)~\cite{borger:asm} into a single framework for capturing transformations within and between modeling languages~\cite{scp-2007}. A transformation is defined by an ASM machine that may contain ASM rules (executable command sequences), graph patterns, GT rules, as well as ASM functions for temporary storage. An optional main rule can serve as entry point. For model manipulation and pattern matching, the transformation may rely on the metamodels available in the VPM model space; such references are made easier by namespace imports. 

\emph{Transformation Execution.}
Transformations are executed within the framework  by using the \Viatra\ interpreter. For pattern matching both (i) \emph{local search based pattern matching} (LS) and (ii) \emph{incremental pattern matching} (INC) are available. This feature provides the transformation designer additional opportunities to finetune the transformation either for faster execution (INC) or lower memory consumption (LS)~\cite{STTT09:AntWorld}.



\section{Transformation tasks} \label{sec:solution}


The \Viatra{} framework has been applied to the subtasks of the \emph{Hello World!} case~\cite{helloworldcase} using the \Viatra{} Textual Command Language (VTCL)~\cite{sac06_vtcl}. Since the case was intended to provide beginners with instructive transformations, we decided to show as many features of the \Viatra{} framework as possible, while also keeping each example as simple as possible. Therefore, each task was solved by multiple variants, which may share pattern definitions, but are otherwise different from each other. The differences are explained at the description of each task.

\subsection{Hello World!} This task included very basic model construction, model-to-model and model-to-text transformations. Since VTCL combines two formalisms for specifying graph transformations, we implemented the different parts of the task with both declarative ASM rules (see~\autoref{lst:helloASM}) and graph transformation rules (see~\autoref{lst:helloGT}). The difference is how much of the model manipulation is described declaratively by GT rules (composed of a LHS and RHS pattern, and potentially an additional ASM action), as opposed to elementary manipulation operations issued in ASM rules; see the homepage\footnote{ \url{http://wiki.eclipse.org/VIATRA2/UseCases/TransformationBestPractices}} for advice on choosing between the two alternative approaches in practice. Although the task seems almost trivial, the accompanying transformation definition is not especially short. This may be taken as a disadvantage of VTCL, but we would like to note, that the we feel that the verbose self-descriptive nature of VTCL aids in comprehension. 

\subsection{Count Matches with certain Properties} This task included parts, where the number of matches are counted using transformations. In the ASM variant (see~\autoref{lst:countASM}), the counting and matching are clearly separated, since the patterns describe what we look for and the forall construct iterates through the matches to count them.

To demonstrate reusability and modularity in VTCL, each part reuses the same ASM rule for creating the result structure. Additionally, the solution references graph patterns defined externally in a separate VTCL file (see~\autoref{lst:graphPatterns}), which acts as a library of common graph patterns. Several further solutions also reuse these patterns. In each case, the VTCL machine corresponding to the library must be loaded first.

The verbosity, once again, is partly voluntary: many of the statements in the graph patterns of~\autoref{lst:graphPatterns} merely assert the type of nodes, and could be safely omitted thanks to type inference. While such verbosity aids in understanding the graph pattern, it does not neccessarily place any additional burden on the developer, as patterns like these can easily be created by selecting some related elements in the model and then exporting their configuration as a graph pattern. 

The upcoming match counting feature of \Viatra{} is still under development (and currently only partially supported), but it will allow for a more elegant solution (see~\autoref{lst:countMC}) of this task. We expect that this solution (and other transformations using match counting within a graph pattern) will be fully supported in release $3.3$ of \Viatra{}.


\subsection{Reverse Edges} In this task, the edges of the graph had to be reversed by the transformation. As before, we created both an ASM (see~\autoref{lst:revASM}) and a GT rule variant (see~\autoref{lst:revGT}). However, here the ASM variant shows an interesting feature of \Viatra{}, as the relations are not modified, only their type is changed from \emph{src} to \emph{trg}, and the other way around. Furthermore, we implemented a third variant (see~\autoref{lst:revRel}) that reverses the edges by switching the target of the \emph{src} relation with the target of the \emph{trg} relation.

Note that using the appropriate conditional language elements (\code{if}, \code{try}), our solutions are tolerant of dangling edges. By ignoring this possibility, the transformation could have been made somewhat simpler.

\subsection{Simple Migration} In this task, the input graph is transformed to a graph conforming to another metamodel. As the case description did not specify it, we implemented both a copy (see~\autoref{lst:migr} and~\autoref{lst:migrTop}) and an in-place (retyping) variant (\autoref{lst:migrInp} and \autoref{lst:migrTopInp}) for both the core and the topology changing transformation. The copy variants simply create the graph using the other metamodel, while the in-place variants use the above mentioned feature of \Viatra{} and change the type of the elements without modifying the rest of the model.

\subsection{Delete Node with Specific Name and its Incident Edges} This task included delete transformations for one specific node and it’s incident edges. We implemented an ASM (\autoref{lst:delASM} and \autoref{lst:delIncASM}) and a GT variant (\autoref{lst:delGT} and \autoref{lst:delIncGT}) for both the core task and the optional task.

\subsection{Insert Transitive Edges} Finally, the last task dealt with inserting transitive edges between nodes. In this case we provide three versions, each with two implementations using ASM rules and GT rules.
\begin{itemize}
  \item First, closely following the original problem specification, we insert edges
between nodes that are 2-hop reachable through an inner node (see \autoref{lst:transASM}
and \autoref{lst:transGT}), i.e. if there are two nodes that are not connected directly, but through an intermediate node, we establish a direct connection between them.
  \item In the next version, we iterate this step as long as applicable so that eventually all transitive reachability edges are inserted (see \autoref{lst:transIterASM} and \autoref{lst:transIterGT}). Note, that although there is only a slight difference in the code of the first two versions, they are independent in implementation.
  \item Finally, we present a solution where all missing transitive reachability edges are detected and inserted in a single step (see \autoref{lst:transAllASM} and
\autoref{lst:transAllGT}). The transformation relies on a graph pattern that expresses full transitive reachability in the graph, using the recursive pattern definition feature of \Viatra{}. One could even argue that actually inserting the missing reachability edges is not necessary in many cases if such pattern matching capability is at our disposal.  Note that due to the recursive nature of the pattern, the current version of the incremental pattern matcher would not work correctly (in case of deleting edges from graphs containing cycles), therefore the default local search-based graph pattern matcher is used. For the other tasks and solution variants the incremental pattern matcher is used.
\end{itemize}

\section{Conclusion}\label{sec:conclusion}
In the current paper we have presented our \Viatra{} based implementation for the Hello World! case study~\cite{helloworldcase}. 

The high points of our solution are (i) the different variants that are self-descriptive and instructive, (ii) the reusable patterns, (iii) the support for recursive matching and (iv) match counting. Furthermore, the dynamic type modification (metamodel manipulation in general) is a highly usable feature especially for migration problems.

On the other hand, since \Viatra{} does not handle EMF models natively, importing and exporting of models is required. Furthermore, as the transformation language is quite verbose, transformations may appear more complex than they really are (note however, that conciseness was not our primary goal when creating these instructive examples). 

Our overall impression is that this simple case study is an excellent basis of comparison of various model transformation tools, filling a long-standing gap.  

\bibliographystyle{eptcs}
\bibliography{bib/ttc10}

\appendix\newpage
\section{Solution demo and implementation}

The deployable implementation and source code is available as an Eclipse online update site (\url{http://mit.bme.hu/~ujhelyiz/viatra/ttc11/}) and the project including the transformations as an archive (\url{http://mit.bme.hu/~ujhelyiz/viatra/ttc11-helloworld.zip})

The SHARE image~\cite{share:viatra} usable for demonstration purposes 
 contains our solution for both the Hello World! and Program Understanding cases.

\section{Appendix - Hello World! transformations}
\label{app:xform}
\subsection{Hello World!}

\lstset{escapeinside={(*}{*)}}
\lstset{numbers=left,stepnumber=10,firstnumber=1}

\begin{lstlisting}[label=lst:helloASM,caption={Hello World transformation, ASM variant},numberfirstline=false]
import datatypes; // imported parts of the model-space are usable by local name  (*\label{asm:label}*)
import nemf.packages;
import nemf.ecore.datatypes;

@incremental // uses incremental pattern-matcher
machine helloWorldASM{
  rule (*\callAnchor{main}*)() = seq{
    println("2.1 Hello World transformation started");

    println("Creating Simple Model with ASM Rule");
    call (*\callLink{createSimpleModelInstance}*)(); // invokes the ASM Rule
    let Greeting = undef in seq{ // define local variable
      println("Creating Extended Model with ASM Rule");
      call (*\callLink{createExtendedModelInstance}*)(Greeting);
      println("Executing model-to-text with ASM Rule");
      call (*\callLink{outputGreeting}*)(Greeting);
    }

    println("2.1 Hello World transformation finished");
  }

  // ASM Rule variant of simple Hello World model instance creation
  rule (*\callAnchor{createSimpleModelInstance}*)() =
   let Greeting = undef, Text = undef, TextRelation = undef in seq{
    // entity creation with explicit parent (in)
    new(helloworld.Greeting(Greeting) in nemf.resources);

    new(EString(Text) in Greeting);
    // setting entity value to some primitive datatype value
    setValue(Text,"Hello world");
    // create relation between elements
    new(helloworld.Greeting.text(TextRelation,Greeting,Text));
  }

  // ASM Rule variant of extended Hello World model instance creation
  rule (*\callAnchor{createExtendedModelInstance}*)(out Greeting) =
   let GreetingMessage = undef, GreetingMessageRelation = undef,
   Text = undef, TextRelation = undef, Person = undef,
    PersonRelation = undef, Name = undef, NameRelation = undef in seq{
    new(helloworldext.Greeting(Greeting) in nemf.resources);

    new(helloworldext.GreetingMessage(GreetingMessage) in Greeting);
    new(helloworldext.Greeting.greetingMessage(GreetingMessageRelation,
     Greeting,GreetingMessage));
    new(EString(Text) in GreetingMessage);
    setValue(Text,"Hello");
    new(helloworldext.GreetingMessage.text(TextRelation,GreetingMessage,Text));

    new(helloworldext.Person(Person) in Greeting);
    new(helloworldext.Greeting.person(PersonRelation,Greeting,Person));
    new(EString(Name) in Person);
    setValue(Name,"TTC Participants");
    new(helloworldext.Person.name(NameRelation,Person,Name));
  }

  // ASM Rule variant of model-to-text transformation
  rule (*\callAnchor{outputGreeting}*)(in Greeting) = let Output = undef, ResR = undef,
   Result = undef in seq{
    /* parameters of "choose" are set by the patternmatcher
     based on matches to the patterns after "find" */
    try choose GreetingMessageText,PersonName with
     find (*\callLink{TextAndNameForGreeting}*)(Greeting,GreetingMessageText,PersonName) do seq{
      new(result.StringResult(Output) in nemf.resources);
      new(EString(Result) in Output);
      new(result.StringResult.result(ResR,Output,Result));
      // value can be set baseed on values from other elements
      setValue(Result,(value(GreetingMessageText) + " " + value(PersonName) + "!"));
    }
  }

  // finds (or creates) Greeting, GreetingMessage.Text and Person.Name
  pattern (*\callAnchor{TextAndNameForGreeting}*)(Greeting,Text,Name) = {

    helloworldext.Greeting(Greeting) in nemf.resources;

    helloworldext.GreetingMessage(GreetingMessage);
    helloworldext.Greeting.greetingMessage(GreetingMessageRelation,
     Greeting,GreetingMessage);
    EString(Text);
    helloworldext.GreetingMessage.text(TextRelation,GreetingMessage,Text);

    helloworldext.Person(Person);
    helloworldext.Greeting.person(PersonRelation,Greeting,Person);
    EString(Name);
    helloworldext.Person.name(NameRelation,Person,Name);
  }
}
\end{lstlisting}

\begin{lstlisting}[label=lst:helloGT,caption={Hello World transformation, GT variant},numberfirstline=false]
import datatypes; // imported parts of the model-space are usable by local name
import nemf.packages;
import nemf.ecore.datatypes;

@incremental // uses incremental patternmatcher
machine helloWorldGT{
  rule main() = seq{
    println("2.1 Hello World transformation started");

    /* "choose" executes once or fails if it cannot, the "try" keyword will let
     the transformation continue even if the "choose" fails */
    try choose with apply (*\callLink{createSimpleModelInstanctGT}*)()
     do println("Creating Simple Model with ASM Rule");
    let Greeting = undef in seq{
      try choose with apply (*\callLink{createExtendedModelInstanctGT}*)(Greeting)
       do println("Creating Extended Model with ASM Rule");
      println("Executing model-to-text with ASM Rule");
      try choose with apply (*\callLink{outputGreetingGT}*)(Greeting) do skip;
    }
    println("2.1 Hello World transformation finished");
  }

  // finds (or creates) Greeting, GreetingMessage.Text and Person.Name
  pattern (*\callQualifiedAnchor{GT.TextAndNameForGreeting}{TextAndNameForGreeting}*)(Greeting,Text,Name) = {

    helloworldext.Greeting(Greeting) in nemf.resources;

    helloworldext.GreetingMessage(GreetingMessage) in Greeting;
    helloworldext.Greeting.greetingMessage(GreetingMessageRelation,
     Greeting,GreetingMessage);
    EString(Text) in GreetingMessage;
    helloworldext.GreetingMessage.text(TextRelation,GreetingMessage,Text);

    helloworldext.Person(Person) in Greeting;
    helloworldext.Greeting.person(PersonRelation,Greeting,Person);
    EString(Name) in Person;
    helloworldext.Person.name(NameRelation,Person,Name);
  }

  // GT Rule variant of simple Hello World model instance creation
  gtrule (*\callAnchor{createSimpleModelInstanctGT}*)() = {
    // the "precondition" is true before the application of the GT Rule
    precondition pattern empty()= {
      // negative application condition (must not match)
      neg pattern (*\callAnchor{existsGreeting}*)(Greeting) = {
        helloworld.Greeting(Greeting);
      }
    }
    // the "postcondition" is true after the application of the GT Rule
    postcondition pattern (*\callAnchor{createdGreeting}*)(Text) = {
      helloworld.Greeting(Greeting) in nemf.resources;
      EString(Text) in Greeting;
      helloworld.Greeting.text(TextRelation,Greeting,Text);
    }
    action { // additional ASM based manipulations after GT Rule application
      setValue(Text,"Hello world");
    }
  }

  // GT Rule variant of extended Hello World model instance creation
  gtrule (*\callAnchor{createExtendedModelInstanctGT}*)(out Greeting) = {
    precondition pattern empty()= {
      neg pattern existsGreeting(Greeting) = {
        helloworldext.Greeting(Greeting);
      }
    }
    postcondition find (*\callQualifiedLink{GT.TextAndNameForGreeting}{TextAndNameForGreeting}*)(Greeting, Text, Name)

    action {
      setValue(Text,"Hello");
      setValue(Name,"TTC Participants");
    }
  }

  // GT Rule variant of model-to-text transformation
  gtrule (*\callAnchor{outputGreetingGT}*)(in Greeting) = {
    precondition find (*\callLink{TextAndNameForGreeting}*)(Greeting, GreetingMessageText, PersonName)

    postcondition pattern (*\callAnchor{outputString}*)(Result) = {
      result.StringResult(Output) in nemf.resources;
      EString(Result) in Output;
      result.StringResult.result(ResR,Output,Result);
    }
    action{
      setValue(Result, value(GreetingMessageText) + " " + value(PersonName) + "!");
    }
  }
}
\end{lstlisting}

\subsection{Common patterns for the Graph metamodels}

\begin{lstlisting}[label=lst:graphPatterns,caption={Common graph patterns},numberfirstline=false]
import datatypes;
import nemf.packages;
import nemf.ecore.datatypes;

machine graphPatterns
{
  // simple type wrapper for Graph
  pattern (*\callQualifiedAnchor{graphPatterns.Graph}{Graph}*)(Graph) = {
    graph1.Graph(Graph);
  }

  // simple type wrapper for Node
  pattern (*\callQualifiedAnchor{graphPatterns.SimpleNode}{SimpleNode}*)(Node) = {
    graph1.Node(Node);
  }

  // finds the nodes and name relation of a node
  pattern (*\callQualifiedAnchor{graphPatterns.NodesRelations}{NodesRelations}*)(Graph,Node,NodesRelation,NameRelation) = {
    graph1.Node(Node);
    graph1.Graph(Graph);
    graph1.Graph.nodes(NodesRelation,Graph,Node);
    EString(Name);
    graph1.Node.name(NameRelation,Node,Name);
  }

  // finds name from the name relation of a node
  pattern (*\callQualifiedAnchor{graphPatterns.nameOfNode}{nameOfNode}*)(NameRelation,Name) = {
    graph1.Node(Node);
    EString(Name);
    graph1.Node.name(NameRelation,Node,Name);
  }

  // simple type wrapper for Edge
  pattern (*\callQualifiedAnchor{graphPatterns.Edge}{Edge}*)(Edge) = {
    graph1.Edge(Edge);
  }

  // simple type wrapper for Edge in Graph
  pattern (*\callQualifiedAnchor{graphPatterns.EdgeOfGraph}{EdgeOfGraph}*)(Graph,Edge) = {
    graph1.Edge(Edge);
    graph1.Graph(Graph);
    graph1.Graph.edges(EdgesRelation,Graph,Edge);
  }

  // finds the edges relation for a node
  pattern (*\callQualifiedAnchor{graphPatterns.EdgesRelation}{EdgesRelation}*)(Graph,Edge,EdgesRelation) = {
    graph1.Edge(Edge);
    graph1.Graph(Graph);
    graph1.Graph.edges(EdgesRelation,Graph,Edge);
  }

  // finds src relation for Edge
  pattern (*\callQualifiedAnchor{graphPatterns.srcAndRelForEdge}{srcAndRelForEdge}*)(Edge,From,SourceRelation) = {
    graph1.Node(From);
    graph1.Edge(Edge);
    graph1.Edge.src(SourceRelation,Edge,From);
  }

  // finds trg relation for Edge
  pattern (*\callQualifiedAnchor{graphPatterns.trgAndRelForEdge}{trgAndRelForEdge}*)(Edge,To,TargetRelation) = {
    graph1.Node(To);
    graph1.Edge(Edge);
    graph1.Edge.trg(TargetRelation,Edge,To);
  }

  // finds looping edges
  pattern (*\callQualifiedAnchor{graphPatterns.loopingEdge}{loopingEdge}*)(Edge) = {
    find (*\callQualifiedLink{graphPatterns.edgeFromToInternal}{edgeFromToInternal}*)(Edge,Node,Node);
  }

  // From is connected with an edge To
  shareable pattern (*\callQualifiedAnchor{graphPatterns.edgeFromToInternal}{edgeFromToInternal}*)(Edge,From,To) = {
    graph1.Node(From);
    graph1.Node(To);
    find (*\callQualifiedLink{graphPatterns.srcAndRelForEdge}{srcAndRelForEdge}*)(Edge,From,SourceRelation);
    find (*\callQualifiedLink{graphPatterns.trgAndRelForEdge}{trgAndRelForEdge}*)(Edge,To,TargetRelation);
  }

  // From is connected with an edge To
  shareable pattern (*\callQualifiedAnchor{graphPatterns.edgeFromTo}{edgeFromTo}*)(From,To) = {
    find (*\callQualifiedLink{graphPatterns.edgeFromToInternal}{edgeFromToInternal}*)(Edge,From,To);
  }

  // From is connected with an edge To and both in Graph
  pattern (*\callQualifiedAnchor{graphPatterns.edgeFromToInGraph}{edgeFromToInGraph}*)(From,To,Graph) = {
    find (*\callQualifiedLink{graphPatterns.edgeFromToInternal}{edgeFromToInternal}*)(Edge,From,To);
    graph1.Graph(Graph);
    graph1.Graph.edges(EdgesRelation,Graph,Edge);
  }


  // finds isolated nodes
  pattern (*\callQualifiedAnchor{graphPatterns.isolatedNode}{isolatedNode}*)(Node) = {
    graph1.Node(Node);
    neg find (*\callQualifiedLink{graphPatterns.srcAndRelForEdge}{srcAndRelForEdge}*)(Edge,Node,SourceRelation); // is not a source
    neg find (*\callQualifiedLink{graphPatterns.trgAndRelForEdge}{trgAndRelForEdge}*)(Edge,Node,TargetRelation); // is not a target
  }

  // three node in a circle
  pattern (*\callQualifiedAnchor{graphPatterns.circleOfThreeNode}{circleOfThreeNode}*)(Node,Inner1,Inner2) = {
    graph1.Node(Node);
    find (*\callQualifiedLink{graphPatterns.edgeFromTo}{edgeFromTo}*)(Node,Inner1);
    find (*\callQualifiedLink{graphPatterns.edgeFromTo}{edgeFromTo}*)(Inner1,Inner2);
    find (*\callQualifiedLink{graphPatterns.edgeFromTo}{edgeFromTo}*)(Inner2,Node);
  }

  // edge with either source or target missing
  pattern (*\callQualifiedAnchor{graphPatterns.danglingEdge}{danglingEdge}*)(Edge) = {// has source but no target
    find (*\callQualifiedLink{graphPatterns.srcAndRelForEdge}{srcAndRelForEdge}*)(Edge,From,SourceRelation);
    neg find (*\callQualifiedLink{graphPatterns.trgAndRelForEdge}{trgAndRelForEdge}*)(Edge,To,TargetRelation);
  } or { // has target but no source
    find (*\callQualifiedLink{graphPatterns.trgAndRelForEdge}{trgAndRelForEdge}*)(Edge,To,TargetRelation);
    neg find (*\callQualifiedLink{graphPatterns.srcAndRelForEdge}{srcAndRelForEdge}*)(Edge,From,SourceRelation);
  }

  // finds the Source of Edge and the corresponding node in the evolved model
  pattern (*\callQualifiedAnchor{graphPatterns.OldAndNewSourceOfEdge}{OldAndNewSourceOfEdge}*)(Edge,Source,Node2) = {
    find (*\callQualifiedLink{graphPatterns.srcAndRelForEdge}{srcAndRelForEdge}*)(Edge,Source,SourceRelation);
    graph2.Node(Node2);
    relation(Traceability,Source,Node2);
  }

  // finds the Target of Edge and the corresponding node in the evolved model
  pattern (*\callQualifiedAnchor{graphPatterns.OldAndNewTargetOfEdge}{OldAndNewTargetOfEdge}*)(Edge,Target,Node2) = {
    find (*\callQualifiedLink{graphPatterns.trgAndRelForEdge}{trgAndRelForEdge}*)(Edge,Target,TargetRelation);
    graph2.Node(Node2);
    relation(Traceability,Target,Node2);
  }

  // finds traceability relations between nodes
  pattern (*\callQualifiedAnchor{graphPatterns.TraceabilityRelation}{TraceabilityRelation}*)(Traceability) = {
    graph1.Node(Node);
    graph2.Node(Node2);
    relation(Traceability,Node,Node2);
  }

  // finds From and To of an edge and the corresponding new nodes
  shareable pattern (*\callQualifiedAnchor{graphPatterns.oldAndNewEdgeFromTo}{oldAndNewEdgeFromTo}*)(From,NewFrom,To,NewTo) = {
    find (*\callQualifiedLink{graphPatterns.edgeFromTo}{edgeFromTo}*)(From,To);
    graph1.Node(From);
    graph3.Node(NewFrom);
    graph1.Node(To);
    graph3.Node(NewTo);
    relation(Tr1,From,NewFrom);
    relation(Tr2,To,NewTo);
  }

  // finds N1 node
  pattern (*\callQualifiedAnchor{graphPatterns.N1Node}{N1Node}*)(Node) = {
    graph1.Node(Node);
    EString(Name);
    graph1.Node.name(NameRel,Node,Name);
    check(value(Name) == "n1");
  }

  // Edge is connected to Node
  pattern (*\callQualifiedAnchor{graphPatterns.connectedEdge}{connectedEdge}*)(Node,Edge) = {
    find (*\callQualifiedLink{graphPatterns.srcAndRelForEdge}{srcAndRelForEdge}*)(Edge,Node,SourceRelation);
  } or {
    find (*\callQualifiedLink{graphPatterns.trgAndRelForEdge}{trgAndRelForEdge}*)(Edge,Node,TargetRelation);
  }

  // From and To (in Graph) are 2-hop transitively connected but not explicitly
  pattern (*\callQualifiedAnchor{graphPatterns.transitiveEdgeMissing2hop}{transitiveEdgeMissing2hop}*)(From,To,Graph) = {
    find (*\callQualifiedLink{graphPatterns.edgeFromToInGraph}{edgeFromToInGraph}*)(From,Inner,Graph);
    find (*\callQualifiedLink{graphPatterns.edgeFromToInGraph}{edgeFromToInGraph}*)(Inner,To,Graph);
    neg find (*\callQualifiedLink{graphPatterns.edgeFromToInGraph}{edgeFromToInGraph}*)(From,To,Graph);
  }
  
  // From and To (in Graph) are transitively connected but not explicitly
  @localsearch
  pattern (*\callQualifiedAnchor{graphPatterns.transitiveEdgeMissing}{transitiveEdgeMissing}*)(From,To,Graph) = {
    find (*\callQualifiedLink{graphPatterns.transitiveConnected}{transitiveConnected}*)(From,To,Graph);
    neg find (*\callQualifiedLink{graphPatterns.edgeFromToInGraph}{edgeFromToInGraph}*)(From,To,Graph);
  }

  // From and To (in Graph) are transitively connected
  @localsearch
  pattern (*\callQualifiedAnchor{graphPatterns.transitiveConnected}{transitiveConnected}*)(From,To,Graph) = {
    find (*\callQualifiedLink{graphPatterns.edgeFromToInGraph}{edgeFromToInGraph}*)(From,Inner,Graph);
    find (*\callQualifiedLink{graphPatterns.edgeFromToInGraph}{edgeFromToInGraph}*)(Inner,To,Graph);
  } or {
    find (*\callQualifiedLink{graphPatterns.edgeFromToInGraph}{edgeFromToInGraph}*)(From,Inner,Graph);
    find (*\callQualifiedLink{graphPatterns.transitiveConnected}{transitiveConnected}*)(Inner,To,Graph);
  }
}
\end{lstlisting}

\subsection{Count Matches with certain Properties}

\begin{lstlisting}[label=lst:countASM,caption={Count matches transformation, ASM variant},numberfirstline=false]
import datatypes;
import nemf.packages;
import nemf.ecore.datatypes;

@incremental
machine countMatchesASM{

  rule main() = seq{
    println("2.2 Count matches transformation started");
    println("Counting number of nodes with ASM Rule");
    call (*\callLink{countNodes}*)();
    println("Counting looping edges with ASM Rule");
    call (*\callLink{countLoopingEdges}*)();
    println("Counting isolated nodes with ASM Rule");
    call (*\callLink{countIsolatedNodes}*)();
    println("Counting circles of three with ASM Rule");
    call (*\callLink{countCirclesOfThree}*)();
    println("Counting dangling edges with ASM Rule");
    call (*\callLink{countDanglingEdges}*)();
    println("2.2 Count matches transformation finished");

  }

  // ASM Rule variant of simple node counting
  rule (*\callAnchor{countNodes}*)() = let Count = 0 in seq {
    // "forall" is executed on each match of the patterns after "find"
    forall Node with find (*\callLink{graphPatterns.SimpleNode}*)(Node) do seq{
      update Count = Count+1; // update overwrites a variable
    }
    // creates EMF model for result
    call (*\callLink{createResult}*)(Count, "Number of nodes");
  }

  // ASM Rule variant of looping edge counting
  rule (*\callAnchor{countLoopingEdges}*)() = let Count = 0 in seq {

    forall Edge with find (*\callLink{graphPatterns.loopingEdge}*)(Edge) do seq{
      update Count = Count+1;
    }
    call (*\callLink{createResult}*)(Count, "Number of looping edges");
  }

  // ASM Rule variant of isolated node counting
  rule (*\callAnchor{countIsolatedNodes}*)() = let Count = 0 in seq {

    forall Node with find (*\callLink{graphPatterns.isolatedNode}*)(Node) do seq{
      update Count = Count+1;
    }
    call (*\callLink{createResult}*)(Count, "Number of isolated nodes");
  }

  // ASM Rule variant of circle of three counting
  rule (*\callAnchor{countCirclesOfThree}*)() = let Count = 0 in seq {

    forall Node,Inner1,Inner2 with
     find (*\callLink{graphPatterns.circleOfThreeNode}*)(Node,Inner1,Inner2) do seq{
      update Count = Count+1;
    }
    call (*\callLink{createResult}*)(Count, "Number of nodes in circles of three");
  }

  // ASM Rule variant of dangling edge counting
  rule (*\callAnchor{countDanglingEdges}*)() = let Count = 0 in seq {

    forall Edge with find (*\callLink{graphPatterns.danglingEdge}*)(Edge) do seq{
      update Count = Count+1;
    }
    call (*\callLink{createResult}*)(Count, "Number of dangling edges");
  }

  // ASM Rule for result storing
  rule (*\callAnchor{createResult}*)(in ResultValue, in Name) = let Result = undef,
   Value = undef, ResR = undef in seq{
    new(result.IntResult(Result) in nemf.resources);
    new(EInt(Value) in Result);
    new(result.IntResult.result(ResR,Result,Value));
    rename(Value,Name);
    setValue(Value, ResultValue);
  }
}
\end{lstlisting}

\begin{lstlisting}[label=lst:countMC,caption={Count matches transformation, match count variant}, numberfirstline=false]
import datatypes;
import nemf.packages;
import nemf.ecore.datatypes;

@incremental
machine countMatchesMC{

  rule main() = seq{
    println("2.2 Count matches transformation started");
    println("Counting number of nodes with MCRule");
    call (*\callLink{countNodesMC}*)();
    println("Counting looping edges with MC Rule");
    call (*\callLink{countLoopingEdgesMC}*)();
    println("Counting isolated nodes with MC Rule");
    call (*\callLink{countIsolatedNodesMC}*)();
    println("Counting circles of three with MC Rule");
    call (*\callLink{countCirclesOfThreeMC}*)();
    println("Counting dangling edges with MC Rule");
    call (*\callLink{countDanglingEdgesMC}*)();
    println("2.2 Count matches transformation finished");

  }

  pattern (*\callAnchor{countNodesPattern}*)(N) = {
    find (*\callLink{graphPatterns.SimpleNode}*)(Node) # N; // counts the number of nodes
  }

  // MC Rule variant of simple node counting
  rule (*\callAnchor{countNodesMC}*)() = seq {
    try choose Count with find (*\callLink{countNodesPattern}*)(Count) do
     let Result = undef, ResR = undef, NodeCount = undef in seq{
      call (*\callLink{createResult2}*)(Count, "Number of nodes");
    }
  }

  pattern (*\callAnchor{countLoopingEdgesPattern}*)(N) = {
    find (*\callLink{graphPatterns.loopingEdge}*)(Edge) # N;
  }

  // MC Rule variant of looping edge counting
  rule (*\callAnchor{countLoopingEdgesMC}*)() = seq {

    try choose Count with find (*\callLink{countLoopingEdgesPattern}*)(Count) do
     let Result = undef, ResR = undef, LoopCount = undef in seq{
      call (*\callLink{createResult2}*)(Count, "Number of looping edges");
    }
  }

  pattern (*\callAnchor{countIsolatedNodesPattern}*)(N) = {
    find (*\callLink{graphPatterns.isolatedNode}*)(Node) # N;
  }

  // MC Rule variant of isolated node counting
  rule (*\callAnchor{countIsolatedNodesMC}*)() = seq {

    try choose Count with find (*\callLink{countIsolatedNodesPattern}*)(Count) do
     let Result = undef, ResR = undef, IsolatedCount = undef in seq{
      call (*\callLink{createResult2}*)(Count, "Number of isolated nodes");
    }
  }

  pattern (*\callAnchor{countCirclesOfThreePattern}*)(N) = {
    find (*\callLink{graphPatterns.circleOfThreeNode}*)(Node,Inner1,Inner2) # N;
  }

  // MC Rule variant of circles of three counting
  rule (*\callAnchor{countCirclesOfThreeMC}*)() = seq {

    try choose Count with find (*\callLink{countCirclesOfThreePattern}*)(Count) do
     let Result = undef, ResR = undef, CircleCount = undef in seq{
      call (*\callLink{createResult2}*)(Count, "Number of nodes in circles of three");
    }
  }

  pattern (*\callAnchor{countDanglingEdgesPattern}*)(N) = {
    find (*\callLink{graphPatterns.danglingEdge}*)(Node) # N;
  }

  // MC Rule variant of dangling edge counting
  rule (*\callAnchor{countDanglingEdgesMC}*)() = seq {

    try choose Count with find (*\callLink{countDanglingEdgesPattern}*)(Count) do
     let Result = undef, ResR = undef, DanglingCount = undef in seq{
      call (*\callLink{createResult2}*)(Count, "Number of dangling edges");
    }
  }

  // ASM Rule for result storing
  rule (*\callAnchor{createResult2}*)(in ResultValue, in Name) = let Result = undef,
   Value = undef, ResR = undef in seq{
    new(result.IntResult(Result) in nemf.resources);
    new(EInt(Value) in Result);
    new(result.IntResult.result(ResR,Result,Value));
    rename(Value,Name);
    setValue(Value, ResultValue);
  }
}
\end{lstlisting}

\subsection{Reverse Edges}

\begin{lstlisting}[label=lst:revASM,caption={Reverse edges transformation, ASM variant},numberfirstline=false]
import datatypes;
import nemf.packages;
import nemf.ecore.datatypes;

@incremental
machine reverseEdgesASM{

  rule main() = seq{
    println("2.3 Reverse edges transformation started");
    call (*\callLink{reverseEdges}*)();
    println("2.3 Reverse edges transformation finished");
  }

  // ASM Rule variant for reverse edges
  rule (*\callAnchor{reverseEdges}*)() = seq{

    forall Edge with find (*\callLink{graphPatterns.Edge}*)(Edge) do let SR = undef, TR = undef in seq{
      // finds src relation if exists
      println(" > Reversing " + name(Edge) + " edge.");
      try choose Source, SourceRelation with
       find (*\callLink{graphPatterns.srcAndRelForEdge}*)(Edge,Source,SourceRelation) do seq{
         update SR = SourceRelation;
      }
      // finds trg relation if exists
      try choose Target, TargetRelation with
       find (*\callLink{graphPatterns.trgAndRelForEdge}*)(Edge,Target,TargetRelation) do seq{
        update TR = TargetRelation;
      }
      if(SR != undef) seq{
        // replace instanceOf relation
        // instanceOf is a relation type, which can be dynamically deleted
        delete(instanceOf(SR,nemf.packages.graph1.Edge.src));
        new(instanceOf(SR,nemf.packages.graph1.Edge.trg)); // and created
      }
      if(TR != undef) seq{
        // replace instanceOf relation
        delete(instanceOf(TR,nemf.packages.graph1.Edge.trg));
        new(instanceOf(TR,nemf.packages.graph1.Edge.src));
      }
    }
  }
}
\end{lstlisting}

\begin{lstlisting}[label=lst:revGT,caption={Reverse edges transformation, GT variant},numberfirstline=false]
import datatypes;
import nemf.packages;
import nemf.ecore.datatypes;

@incremental
machine reverseEdgesGT{

  rule main() = seq{
    println("2.3 Reverse edges transformation started");
    forall Edge with apply (*\callLink{reverseEdgesGT}*)(Edge) do
      println(" > Reversing " + name(Edge) + " edge.");
    println("2.3 Reverse edges transformation finished");
  }

  // GT Rule variant for reverse edges
  // note: add "shareable" keyword before "pattern" to
  // actually reverse looping edges as well
  gtrule (*\callAnchor{reverseEdgesGT}*)(out Edge) = {
    precondition pattern (*\callAnchor{edgeWithRelations}*)(Edge,From,To,SourceRel,TargetRel) = {
      find (*\callLink{graphPatterns.srcAndRelForEdge}*)(Edge,From,SourceRel);
      find (*\callLink{graphPatterns.trgAndRelForEdge}*)(Edge,To,TargetRel);
    }
    postcondition pattern (*\callAnchor{reversedEdges}*)(Edge,From,To,SourceRel,TargetRel) = {
      find (*\callLink{graphPatterns.srcAndRelForEdge}*)(Edge,To,SourceRel);
      find (*\callLink{graphPatterns.trgAndRelForEdge}*)(Edge,From,TargetRel);
    }
  }
}
\end{lstlisting}

\begin{lstlisting}[label=lst:revRel,caption={Reverse edges transformation, relation manipulation variant},numberfirstline=false]
import datatypes;
import nemf.packages;
import nemf.ecore.datatypes;

@incremental
machine reverseEdgesRel{

  rule main() = seq{
    println("2.3 Reverse edges transformation started");
    call (*\callLink{reverseEdges}*)();
    println("2.3 Reverse edges transformation finished");
  }

  // ASM Rule variant for reverse edges
  rule reverseEdges() = seq{

    forall Edge with find (*\callLink{graphPatterns.Edge}*)(Edge) do
     let S = undef, SR = undef, T = undef, TR = undef in seq{
      // finds src relation if exists
      println(" > Reversing " + name(Edge) + " edge.");
      try choose Source,SourceRelation with
       find (*\callLink{graphPatterns.srcAndRelForEdge}*)(Edge,Source,SourceRelation) do seq{
         update S = Source;
         update SR = SourceRelation;
      }
      // finds trg relation if exists
      try choose Target,TargetRelation with
       find (*\callLink{graphPatterns.trgAndRelForEdge}*)(Edge,Target,TargetRelation) do seq{
        update T = Target;
        update TR = TargetRelation;
      }
      if(T != undef) seq{
        println("   > Reversing target to source: " + name(T));
        if(SR != undef) setTo(SR,T); // change target of relation
        else seq{
          delete(TR);
          new(graph1.Edge.src(SR,Edge,T));
        }
      }
      if(S != undef) seq{
        println("   > Reversing source to target: " + name(S));
        if(TR != undef) setTo(TR,S);
        else seq{
          delete(SR);
          new(graph1.Edge.trg(TR,Edge,S));
        }
      }
    }
  }
}
\end{lstlisting}

\subsection{Simple Migration}

\begin{lstlisting}[label=lst:migr,caption={Simple Migration transformation, copy variant},numberfirstline=false]
import datatypes;
import nemf.packages;
import nemf.ecore.datatypes;

@incremental
machine simpleMigration{

  rule main() = seq{

    println("2.4 Simple Migration (with copy) transformation started");
    call (*\callLink{migrateGraph}*)();
    println("2.4 Simple Migration (with copy) transformation finished");
  }

  // ASM Rule variant of simple migration transformation
  rule (*\callAnchor{migrateGraph}*)() = seq{

    forall Graph with find (*\callLink{graphPatterns.Graph}*)(Graph) do
     let Graph2 = undef, GCSRel = undef in seq{
      new(graph2.Graph(Graph2) in nemf.resources); // create graph

      forall Node,NodesRelation, NameRelation with
       // for each node, create a new
       find (*\callLink{graphPatterns.NodesRelations}*)(Graph,Node,NodesRelation, NameRelation) do
       let Node2 = undef, Text = undef, TextRel = undef,Traceability = undef in seq{
        new(graph2.Node(Node2) in Graph2);
        new(graph2.Graph.gcs(GCSRel,Graph2,Node2));
        new(EString(Text) in Node2);
        try choose Name with find (*\callLink{graphPatterns.nameOfNode}*)(NameRelation,Name) do seq{
          setValue(Text,value(Name));
        }
        new(graph2.GraphComponent.text(TextRel,Node2,Text));
        // store the traceability between old and new node
        new(relation(Traceability,Node,Node2));
      }

      // for each edge, create a new
      forall Edge,EdgesRelation with
       find (*\callLink{graphPatterns.EdgesRelation}*)(Graph,Edge,EdgesRelation) do
       let Edge2 = undef,Text = undef,TextRel = undef, EvolvedRel = undef in seq{
        new(graph2.Edge(Edge2) in Graph2);
        new(graph2.Graph.gcs(GCSRel,Graph2,Edge2));
        new(EString(Text) in Edge2);
        new(graph2.GraphComponent.text(TextRel,Edge,Text));
        // each source relation is created to the corresponding node
        forall Source,Node2 with
         find (*\callLink{graphPatterns.OldAndNewSourceOfEdge}*)(Edge,Source,Node2) do seq{
          new(graph2.Edge.src(EvolvedRel,Edge2,Node2));
        }
        forall Target,Node2 with
         find (*\callLink{graphPatterns.OldAndNewTargetOfEdge}*)(Edge,Target,Node2) do seq{
          // each tagret relation is created to the corresponding node
          new(graph2.Edge.trg(EvolvedRel,Edge2,Node2));
        }
      }

      // delete traceability models
      forall Traceability with
       find (*\callLink{graphPatterns.TraceabilityRelation}*)(Traceability) do seq{
        delete(Traceability);
      }
    }
  }
}
\end{lstlisting}

\begin{lstlisting}[label=lst:migrInp,caption={Simple Migration transformation, in-place variant},numberfirstline=false]
import datatypes;
import nemf.packages;
import nemf.ecore.datatypes;

@incremental
machine simpleMigrationInplace{

  rule main() = seq{

    println("2.4 Simple Migration (in-place) transformation started");
    call (*\callLink{migrateGraphInplace}*)();
    println("2.4 Simple Migration (in-place) transformation finished");
  }

  // ASM Rule variant of simple migration in-place transformation
  rule (*\callAnchor{migrateGraphInplace}*)() = seq{
    // at this point, each Graph is transformed
    forall Graph with find (*\callLink{graphPatterns.Graph}*)(Graph) do seq{

      // each node is transformed using instanceOf changing
      forall Node,NodesRelation, NameRelation with
       find (*\callLink{graphPatterns.NodesRelations}*)(Graph,Node,NodesRelation, NameRelation) do seq{
        delete(instanceOf(Node,nemf.packages.graph1.Node));
        new(instanceOf(Node,nemf.packages.graph2.Node));
        delete(instanceOf(NodesRelation,nemf.packages.graph1.Graph.nodes));
        new(instanceOf(NodesRelation,nemf.packages.graph2.Graph.gcs));
        delete(instanceOf(NameRelation,nemf.packages.graph1.Node.name));
        new(instanceOf(NameRelation,nemf.packages.graph2.GraphComponent.text));
      }

      // each edge is transformed using instanceOf changing
      forall Edge,EdgesRelation with
       find (*\callLink{graphPatterns.EdgesRelation}*)(Graph,Edge,EdgesRelation) do
       let Text = undef,TextRel = undef in seq{
        delete(instanceOf(Edge,nemf.packages.graph1.Edge));
        new(instanceOf(Edge,nemf.packages.graph2.Edge));
        delete(instanceOf(EdgesRelation,nemf.packages.graph1.Graph.edges));
        new(instanceOf(EdgesRelation,nemf.packages.graph2.Graph.gcs));
        new(EString(Text) in Edge);
        new(graph2.GraphComponent.text(TextRel,Edge,Text));
      }
      // the graph is transformed
      delete(instanceOf(Graph,nemf.packages.graph1.Graph));
      new(instanceOf(Graph,nemf.packages.graph2.Graph));
    }
  }
}
\end{lstlisting}

\begin{lstlisting}[label=lst:migrTop,caption={Simple Migration Topology changing transformation, copy variant},numberfirstline=false]
import datatypes;
import nemf.packages;
import nemf.ecore.datatypes;

@incremental
machine simpleMigrationTopology{

  rule main() = seq{

    println("2.4 Simple Migration (topoogy change with copy) transformation started");
    call (*\callLink{topologyChange}*)();
    println("2.4 Simple Migration (topoogy change with copy) transformation finished");
  }

  // ASM Rule variant of topology-changing migration
  rule (*\callAnchor{topologyChange}*)() = seq{
    forall Graph with find (*\callLink{graphPatterns.Graph}*)(Graph) do
     let NewGraph = undef, Rel = undef in seq{
      new(graph3.Graph(NewGraph) in nemf.resources);

      forall Node,NodesRelation, NameRelation with
       find (*\callLink{graphPatterns.NodesRelations}*)(Graph,Node,NodesRelation, NameRelation) do
       let NewNode = undef, Text = undef, Traceability = undef in seq{
        new(graph3.Node(NewNode) in NewGraph);
        new(graph3.Graph.nodes(Rel,NewGraph,NewNode));
        new(EString(Text) in NewNode);
        try choose Name with find (*\callLink{graphPatterns.nameOfNode}*)(NameRelation,Name) do seq{
          setValue(Text,value(Name));
        }
        new(graph3.Node.text(Rel,NewNode,Text));
        new(relation(Traceability,Node,NewNode));
      }

      forall From,To,NewFrom,NewTo with
       find (*\callLink{graphPatterns.oldAndNewEdgeFromTo}*)(From,NewFrom,To,NewTo) do
       let LinksToRel = undef in seq{
        new(graph3.Node.linksTo(LinksToRel,NewFrom,NewTo));
      }
      forall Traceability with
       find (*\callLink{graphPatterns.TraceabilityRelation}*)(Traceability) do seq{
        delete(Traceability);
      }
    }
  }
}
\end{lstlisting}

\begin{lstlisting}[label=lst:migrTopInp,caption={Simple Migration Topology changing transformation, in-place variant},numberfirstline=false]
import datatypes;
import nemf.packages;
import nemf.ecore.datatypes;

@incremental
machine simpleMigrationTopologyInplace{

  rule main() = seq{

    println("2.4 Simple Migration (topology change in-place) transformation started");
    call (*\callLink{topologyChangeInplace}*)();
    println("2.4 Simple Migration (topology change in-place) transformation finished");
  }

  // ASM Rule variant of topology-changing in-place migration
  rule (*\callAnchor{topologyChangeInplace}*)() = seq{
    forall Graph with find (*\callLink{graphPatterns.Graph}*)(Graph) do seq{

      forall Node,NodesRelation, NameRelation with
       find (*\callLink{graphPatterns.NodesRelations}*)(Graph,Node,NodesRelation,NameRelation) do
       seq{ //
        delete(instanceOf(Node,nemf.packages.graph1.Node));
        new(instanceOf(Node,nemf.packages.graph3.Node));
        delete(instanceOf(NodesRelation,nemf.packages.graph1.Graph.nodes));
        new(instanceOf(NodesRelation,nemf.packages.graph3.Graph.nodes));
        delete(instanceOf(NameRelation,nemf.packages.graph1.Node.name));
        new(instanceOf(NameRelation, nemf.packages.graph3.Node.text));

        forall To with find (*\callLink{graphPatterns.edgeFromTo}*)(Node,To) do
         let LinksToRel = undef in seq{
          new(graph3.Node.linksTo(LinksToRel,Node,To));
        }
      }

      forall Edge,EdgesRelation with
       find (*\callLink{graphPatterns.EdgesRelation}*)(Graph,Edge,EdgesRelation) do seq{
        delete(Edge);
      }

      delete(instanceOf(Graph,nemf.packages.graph1.Graph));
      new(instanceOf(Graph, nemf.packages.graph3.Graph));
    }
  }

}
\end{lstlisting}

\subsection{Delete Node with Specific Name and its Incident Edges}

\begin{lstlisting}[label=lst:delASM,caption={Delete node transformation, ASM variant},numberfirstline=false]
import datatypes;
import nemf.packages;
import nemf.ecore.datatypes;

@incremental
machine deleteNodeASM{

  rule main() = seq{
    println("2.5 Delete nodes transformation started");
    call (*\callLink{deleteNode}*)();
    println("2.3 Delete nodes transformation finished");
  }

  rule (*\callAnchor{deleteNode}*)() = seq{
    try choose N1 with find (*\callLink{graphPatterns.N1Node}*)(N1) do seq{
      delete(N1);
    }
  }
}
\end{lstlisting}

\begin{lstlisting}[label=lst:delGT,caption={Delete node transformation, GT variant},numberfirstline=false]
import datatypes;
import nemf.packages;
import nemf.ecore.datatypes;

@incremental
machine deleteNodeGT{

  rule main() = seq{
    println("2.5 Delete n1 node transformation (GT) started");
    try choose with apply (*\callLink{deleteNodeGT}*)() do skip;
    println("2.3 Delete n1 node transformation finished");
  }

  gtrule (*\callAnchor{deleteNodeGT}*)() = {
    precondition find (*\callLink{graphPatterns.N1Node}*)(N1)

    postcondition pattern (*\callAnchor{noN1}*)(N1) = {
      neg find (*\callLink{graphPatterns.N1Node}*)(N1);
    }
  }

}
\end{lstlisting}

\begin{lstlisting}[label=lst:delIncASM,caption={Delete node and incident edges transformation, ASM variant},numberfirstline=false]
import datatypes;
import nemf.packages;
import nemf.ecore.datatypes;

@incremental
machine deleteNodeIncidentASM{

  rule main() = seq{
    println("2.5 Delete nodes transformation started");
    println("Deleting incident edges as well");
    call (*\callLink{deleteNodeAndIncidentEdges}*)();
    println("2.3 Delete nodes transformation finished");
  }

  rule (*\callAnchor{deleteNodeAndIncidentEdges}*)() = seq{
    try choose N1 with find (*\callLink{graphPatterns.N1Node}*)(N1) do seq{
      forall Edge with find (*\callLink{graphPatterns.connectedEdge}*)(N1,Edge) do seq{
        delete(Edge);
      }
      delete(N1);
    }
  }
}
\end{lstlisting}

\begin{lstlisting}[label=lst:delIncGT,caption={Delete node and incident edges transformation, GT variant},numberfirstline=false]
import datatypes;
import nemf.packages;
import nemf.ecore.datatypes;

@incremental
machine deleteNodeIncidentGT{

  rule main() = seq{
    println("2.5 Delete n1 node transformation (GT) started");
    println("Deleting incident edges as well");
    try choose with apply (*\callLink{deleteNodeAndIncidentEdgesGT}*)() do skip;
    println("2.3 Delete n1 node transformation finished");
  }

  gtrule (*\callQualifiedAnchor{incident.deleteNodeGT}{deleteNodeGT}*)(in N1) = {
    precondition find (*\callLink{graphPatterns.N1Node}*)(N1)

    postcondition pattern noN1(N1) = {
      neg find (*\callLink{graphPatterns.N1Node}*)(N1);
    }
  }

  gtrule (*\callAnchor{deleteNodeAndIncidentEdgesGT}*)() = {
    precondition find (*\callLink{graphPatterns.N1Node}*)(N1)

    action{
      forall Edge with apply (*\callLink{deleteIncidentEdgesOfNode}*)(N1,Edge) do skip;
      try choose with apply (*\callQualifiedLink{incident.deleteNodeGT}{deleteNodeGT}*)(N1) do skip;
    }
  }

  gtrule (*\callAnchor{deleteIncidentEdgesOfNode}*)(in Node, out Edge) = {
    precondition find (*\callLink{graphPatterns.connectedEdge}*)(Node,Edge)

    postcondition pattern noConnectingEdge(Node,Edge) = {
      graph1.Node(Node);
      neg find (*\callLink{graphPatterns.connectedEdge}*)(Node,Edge);
    }
  }
}
\end{lstlisting}

\subsection{Insert Transitive Edges}

\begin{lstlisting}[label=lst:transASM,caption={Insert transitive edges transformation, ASM variant},numberfirstline=false]
import datatypes;
import nemf.packages;
import nemf.ecore.datatypes;

@incremental
machine transitiveEdgesASM{

  rule main() = seq{
    println("2.6 Transitive edges (R u R^2) transformation (ASM) started");
    call (*\callLink{insertTransitiveEdgesOnce}*)();
    println("2.6 Transitive edges transformation finished");
  }

  // ASM Rule variant for inserting edges
  // between each pair of transitively connected nodes
  rule (*\callAnchor{insertTransitiveEdgesOnce}*)() = seq{
    forall From, To, Graph with
     find (*\callLink{graphPatterns.transitiveEdgeMissing2hop}*)(From,To,Graph) do
     let TransitiveEdge = undef, Rel = undef in seq{
      new(graph1.Edge(TransitiveEdge) in Graph);
      new(graph1.Graph.edges(Rel,Graph,TransitiveEdge));
      new(graph1.Edge.src(Rel,TransitiveEdge,From));
      new(graph1.Edge.trg(Rel,TransitiveEdge,To));
    }
  }

}
\end{lstlisting}

\begin{lstlisting}[label=lst:transGT,caption={Insert transitive edges transformation, GT variant},numberfirstline=false]
import datatypes;
import nemf.packages;
import nemf.ecore.datatypes;

@incremental
machine transitiveEdgesGT{

  rule main() = seq{
    println("2.6 Transitive edges (R u R^2) transformation (GT) started");
    forall From, To with apply (*\callLink{insertTransitiveEdgesOnceGT}*)(From, To) do skip;
    println("2.6 Transitive edges transformation finished");
  }

  // GT Rule for inserting transitive edges between From and To
  gtrule (*\callAnchor{insertTransitiveEdgesOnceGT}*)(out From, out To) = {
    precondition find (*\callLink{graphPatterns.transitiveEdgeMissing2hop}*)(From,To,Graph)

    postcondition find (*\callLink{graphPatterns.edgeFromToInGraph}*)(From,To,Graph)
  }
}
\end{lstlisting}

\begin{lstlisting}[label=lst:transIterASM,caption={Insert all transitive edges iteratively transformation, ASM variant},numberfirstline=false]
import datatypes;
import nemf.packages;
import nemf.ecore.datatypes;

@incremental
machine transitiveEdgesIterativeASM{

  rule main() = seq{
    println("2.6 Transitive edges (R u R^2) transformation (ASM) started");
    println("Insert edges iteratively");
    call (*\callLink{insertTransitiveEdgesIterative}*)();
    println("2.6 Transitive edges transformation finished");
  }

  // ASM Rule variant for inserting edges between each transitively connected nodes
  rule (*\callAnchor{insertTransitiveEdgesIterative}*)() = seq{
    iterate choose From, To, Graph with
     find (*\callLink{graphPatterns.transitiveEdgeMissing2hop}*)(From,To,Graph) do
     let TransitiveEdge = undef, Rel = undef in seq{
      new(graph1.Edge(TransitiveEdge) in Graph);
      new(graph1.Graph.edges(Rel,Graph,TransitiveEdge));
      new(graph1.Edge.src(Rel,TransitiveEdge,From));
      new(graph1.Edge.trg(Rel,TransitiveEdge,To));
    }
  }

}
\end{lstlisting}

\begin{lstlisting}[label=lst:transIterGT,caption={Insert all transitive edges iteratively transformation, GT variant},numberfirstline=false]
import datatypes;
import nemf.packages;
import nemf.ecore.datatypes;

@incremental
machine transitiveEdgesIterativeGT{

  rule main() = seq{
    println("2.6 Transitive edges (R u R^2) transformation (GT) started");
    println("Insert edges iteratively");
    iterate choose From, To with apply (*\callQualifiedLink{iterate.insertTransitiveEdgesOnceGT}{insertTransitiveEdgesOnceGT}*)(From, To) do skip;
    println("2.6 Transitive edges transformation finished");
  }

  // GT Rule for inserting transitive edges between From and To
  gtrule (*\callQualifiedAnchor{iterate.insertTransitiveEdgesOnceGT}{insertTransitiveEdgesOnceGT}*)(out From, out To) = {
    precondition find (*\callLink{graphPatterns.transitiveEdgeMissing2hop}*)(From,To,Graph)

    postcondition find (*\callLink{graphPatterns.edgeFromToInGraph}*)(From,To,Graph)
  }
}
\end{lstlisting}

\begin{lstlisting}[label=lst:transAllASM,caption={Insert all transitive edges transformation, ASM variant},numberfirstline=false]
import datatypes;
import nemf.packages;
import nemf.ecore.datatypes;

machine transitiveEdgesAllASM{

  rule main() = seq{
    println("2.6 Transitive edges (R u R^2 ... u R^n) transformation (ASM) started");
    call (*\callLink{insertTransitiveEdgesAll}*)();
    println("2.6 Transitive edges transformation finished");
  }

  // ASM Rule variant for inserting edges
  // between each pair of transitively connected nodes
  rule (*\callAnchor{insertTransitiveEdgesAll}*)() = seq{
    forall From, To, Graph with
     find (*\callLink{graphPatterns.transitiveEdgeMissing}*)(From,To,Graph) do
     let TransitiveEdge = undef, Rel = undef in seq{
      new(graph1.Edge(TransitiveEdge) in Graph);
      new(graph1.Graph.edges(Rel,Graph,TransitiveEdge));
      new(graph1.Edge.src(Rel,TransitiveEdge,From));
      new(graph1.Edge.trg(Rel,TransitiveEdge,To));
    }
  }
}
\end{lstlisting}

\begin{lstlisting}[label=lst:transAllGT,caption={Insert all transitive edges transformation, GT variant},numberfirstline=false]
import datatypes;
import nemf.packages;
import nemf.ecore.datatypes;

machine transitiveEdgesAllGT{

  rule main() = seq{
    println("2.6 Transitive edges (R u R^2 ... u R^n) transformation (GT) started");
    forall From, To with apply (*\callLink{insertTransitiveEdgesAllGT}*)(From, To) do skip;
    println("2.6 Transitive edges transformation finished");
  }

  // GT Rule for inserting transitive edges between From and To
  gtrule (*\callAnchor{insertTransitiveEdgesAllGT}*)(out From, out To) = {
    precondition find (*\callLink{graphPatterns.transitiveEdgeMissing}*)(From,To,Graph)

    postcondition find (*\callLink{graphPatterns.edgeFromToInGraph}*)(From,To,Graph)
  }
}
\end{lstlisting}

\end{document}